\documentclass[10pt,journal,twoside]{IEEEtran}
\usepackage{cite}

\ifCLASSINFOpdf
\usepackage[pdftex]{graphicx}
\else
\fi

\begin{document}
%
\title{High TMR ratio in Co$_2$FeSi and Fe$_2$CoSi based magnetic tunnel junctions}
%
%
%


\author{
\IEEEauthorblockN{Christian~Sterwerf,~Markus~Meinert,~Jan-Michael~Schmalhorst,~and~G\"unter~Reiss}\\[2mm]
\IEEEauthorblockA{\small{Thin Films and Physics of Nanostructures, Department of Physics, Bielefeld University, 33501 Bielefeld, Germany}}
}

\maketitle

%
\begin{abstract}
\boldmath
Magnetic tunnel junctions with Fe$_{1+x}$Co$_{2-x}$Si ($0 \leq x \leq 1$) electrodes and MgO barrier were prepared on MgO substrates by magnetron co-sputtering. Maximum tunnel magnetoresistance (TMR) ratios of $262\%$ at $15$\,K and $159\%$ at room temperature were observed for $x = 0.75$. Correlations of the annealing temperature dependent atomic ordering and TMR amplitude are discussed. The high TMR for an intermediate stoichiometry is ascribed to the adjustment of the Fermi energy within a minority spin pseudo gap.

\end{abstract}


%

\section{Introduction}

%
%
%
%
\IEEEPARstart{S}{pintronic} devices exploit the electron charge and spin for data communication and storage. Commercial applications of spintronic devices include read heads of modern hard drives and non-volatile MRAMs (magnetoresistive random access memory) \cite{Tehrani:1999vq}.
Half-metallic materials are favorable candidates for spintronic applications because of a gap in one spin channel, leading to full spin polarization at the Fermi energy \cite{Pickett01}. Thus, large tunnel magnetoresistance (TMR) ratios can be obtained. This effect is observed in magnetic tunnel junctions, consisting of two ferromagnetic films separated by a thin insulator. The TMR ratio is defined by (R$_{\mathrm{AP}}$-R$_{\mathrm{P}}$)/R$_{\mathrm{P}}$, where R$_{\mathrm{P}}$ and R$_{\mathrm{AP}}$ denote the resistivity in parallel and antiparallel magnetization alignment, respectively. 

Half-metallicity has been predicted for many materials such as La$_{1-x}$Sr$_{x}$MnO$_3$ \cite{Picket}, CrO$_2$ \cite{Coey:2002ud}, and many Heusler compounds \cite{Galanakis:2002ty}. Especially Co based Heusler compounds are suitable for commercial applications because of their high Curie temperatures. A ternary Heusler compound is represented by the formula X$_2$YZ, in which X and Y are transition metals and Z is a main group element. These crystallize in one of two possible fully ordered fcc structures with a four-atom basis: an ordinary (L2$_1$) structure and an inverse structure (X$_a$) with a different occupation sequence \cite{Luo:2007io}. The prototype of the L2$_1$ structure is the original Heusler compound Cu$_2$MnAl (space group Fm$\bar{3}$m) \cite{Bradley1934}. The occupation sequence is in the L2$_1$ case X-Y-X-Z with respect to the sublattices A($\vec{0}$), B($\vec{\frac{1}{4}}$), C($\vec{\frac{1}{2}}$), and D($\vec{\frac{3}{4}}$). The condition for a compound to crystallize in the X$_a$ structure is that element X is less electronegative than element Y. Within one row of the periodic table, this is the case when the atomic number of Y is higher than that of  X \cite{Graf:2011jj}. The prototypes for this structure are Hg$_2$CuTi and Li$_2$AgSb \cite{puselj,pauly}. No inversion symmetry is present and the symmetry is lowered to F$\bar{4}3$m. The occupation of the sublattices is Y-X-X-Z, with the two X-atoms at inequivalent positions in the lattice. Here we study thin films with compositions between Co$_2$FeSi and Fe$_2$CoSi, which crystallize in the regular Heusler and inverse Heusler structures, respectively \cite{Balke06,Luo:2007io}. According to band structure calculations, these materials should have (pseudo-)gaps in the minority states, such that the Fermi energy is close to the flat valence bands in Fe$_2$CoSi \cite{Luo:2007io} and close to the flat conduction bands in Co$_2$FeSi \cite{Gercsi07}. We thus follow the idea of Fermi energy adjustment by electron doping \cite{Kubota09, Sakuraba10, Schmalhorst:2008wz, Gercsi07} to move the Fermi energy away from the flat bands, as to increase the TMR. This is expected in the Fe$_{1+x}$Co$_{2-x}$Si series under the assumption of an ideal substitution of Fe at the C site with Co. Furthermore, the lattice constant is expected to remain constant in this series, so lattice mismatches with the substrate or the tunneling barrier should not vary.

\section{Preparation and Characterization Techniques}
The films were deposited in a dc/rf  magnetron co-sputtering system with a typical base pressure of $1\cdot 10^{-9}$\,mbar. The Ar pressure during sputtering was $2\cdot 10^{-3}$\,mbar. All films were deposited on MgO (001) substrates (lattice parameter: 4.21\,\AA{}) covered by a 5\,nm thick MgO and a 5\,nm thick Cr seed layer. The seed layers were \textit{in situ} annealed at $700^\circ$C to obtain smooth surfaces. Afterwards, epitaxial Fe$_{1+x}$Co$_{2-x}$Si films were grown by co-sputtering from elemental targets at room temperature. The small lattice mismatch between two unit cells of Cr (2$\times$2.88\,\AA{}) and one unit cell of Co$_2$FeSi (5.64\,\AA{} \cite{Wurmehl:2005ia}) or Fe$_2$CoSi (5.645\,\AA{} \cite{Luo:2007io}) allows a coherent growth. The correct stoichiometries were obtained by adjusting the sputtering powers according to x-ray fluorescence measurements. 2\,nm MgO films were sputter deposited as protective layers or as tunnel barrier on top. For the MTJs, counter-electrodes of Co$_{70}$Fe$_{30}$ 5\,nm / Mn$_{83}$Fe$_{17}$ 10\,nm / Ru 25\,nm were finally deposited at room temperature. The tunneling junctions were patterned into (7.5$\times$7.5)\,$\mu$m$^2$, (12.5$\times$12.5)\,$\mu$m$^2$, and (22.5$\times$22.5)\,$\mu$m$^2$ elements by UV lithography and ion beam etching. The layer stacks were subsequently \textit{ex situ} vacuum annealed and the crystallographic properties and the TMR ratios were investigated in dependence of the post-annealing temperature $T_\mathrm{a}$.

The transport measurements were carried out with a two-point probe technique in a magnetic field of up to 0.3\,T. Low-temperature measurements down to 15\,K were performed in a closed-cycle He cryostat. X-ray diffraction was performed with a Philips X'pert Pro MPD diffractometer with a copper anode and Bragg-Brentano arrangement. The chemical disorder was analyzed with off-specular measurements on an open Euler cradle and point focus collimator optics.

\section{Results}

\subsection{Crystal structure}
\begin{figure}[!t]
\centering
\includegraphics[width=1\linewidth]{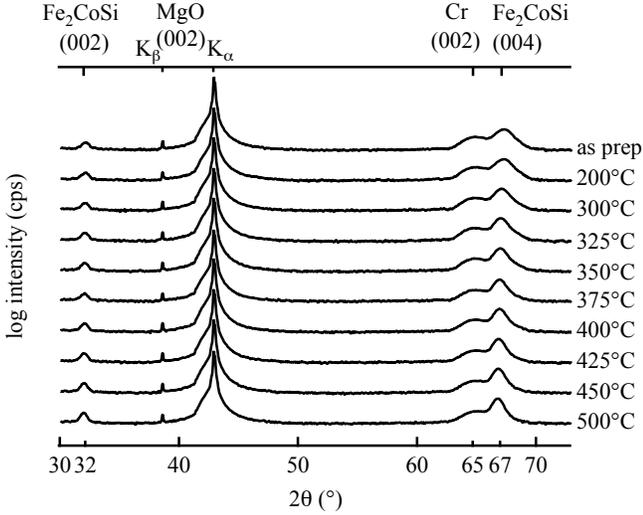}
\caption{Diffractograms of Fe$_2$CoSi films for the as prepared state and different post-annealing temperatures. The curves are shifted for clarity.}
\label{Fe2CoSiXRD}
\end{figure}

XRD patterns for all stoichiometries were taken for the as prepared state and a post-annealing series ranging from $200^\circ$C to $500^\circ$C. Figure \ref{Fe2CoSiXRD} depicts the diffractograms exemplarily for the inverse Heusler compound Fe$_2$CoSi. The diffractograms of Co$_2$FeSi and the intermediate stoichiometries show a very similar behavior. The Cr diffraction peak becomes less intense at annealing temperatures higher than $500^\circ$C due to diffusion into the Heusler compound (not shown).

\begin{figure}[!t]
\centering
\includegraphics[width=1\linewidth]{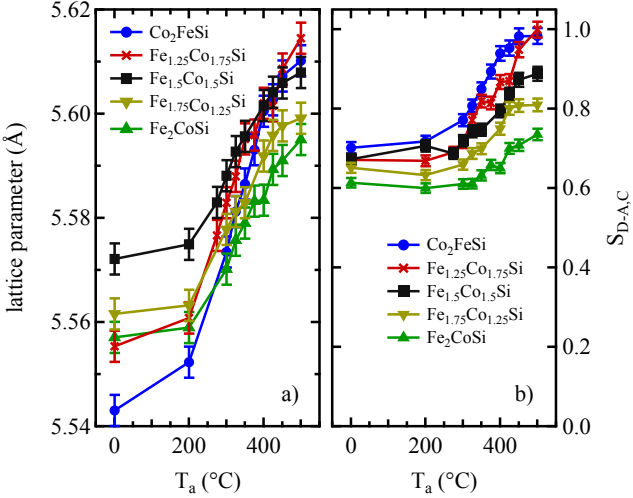}
\caption{Influence of the post-annealing temperature $T_\mathrm{a}$ on a) the lattice parameter and b) $S_{\mathrm{D-A,C}}$ ordering parameter.}
\label{a004}
\end{figure}
Figure \ref{a004} a) shows the $T_\mathrm{a}$ dependence of the lattice parameter for x=0, 0.25, 0.5, 0.75, 1. The lattice parameter increases for all samples towards the bulk value but they do not seem to be completely converged at $500^\circ$C and may further increase at higher temperatures. The lattice constant of Cr is larger than that of the Fe$_{1+x}$Co$_{2-x}$Si films, so a strained state with an enlarged in-plane lattice constant and a compressed out-of-plane lattice constant is likely to occur. This strain is relaxed during post-annealing.

\begin{figure}[!t]
\centering
\includegraphics[width=\linewidth]{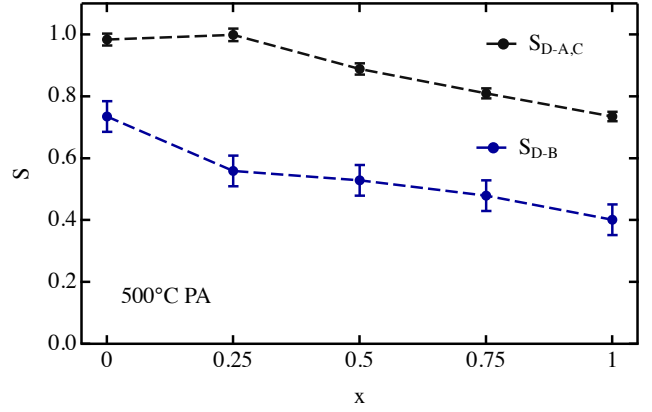}
\caption{$S_{\mathrm{D-A,C}}$ and $S_{\mathrm{D-B}}$ ordering parameters for the Fe$_{1+x}$Co$_{2-x}$Si films post-annealed at 500$^\circ$C.}%
\label{S_vs_x_graph}%
\end{figure}%

Due to the use of Cu\,K$_\alpha$ radiation, we are only able to distinguish the ordering between the Si and the Co or Fe atoms, because the atomic scattering factors of Co and Fe are almost the same for this energy \cite{Henke}. For this reason, we do not see any difference between the ordinary and the inverse Heusler structure in the diffractograms.
To analyze the atomic ordering, we reduce the L2$_1$ and X$_a$ structures to the D0$_3$ structure, where the Co and Fe atoms are taken as equivalent. 
Takamura et al. \cite{Takamura:2009ei} proposed a model for the L2$_1$ and B2 ordering parameters for Heusler compounds with the formula X$_2$YZ. Transferring this model to a D0$_3$ structure, one obtains the same formulas as in Takamura's formalism. For unification of the ordering parameters for the different structures, we denote these parameters with respect to the ordering between the sublattices A, B, C, and D. The degree of ordering between sublattices D (occupied by Si in both ideal cases) and A, C is defined as
\begin{eqnarray}
S_{\mathrm{D-A,C}}=2z-1.
\end{eqnarray}
The ordering between sublattice D and B is determined by the relation:
\begin{eqnarray}
S_{\mathrm{D-B}}=\frac{2y-z}{2-z},
\end{eqnarray}
where $z$ and $y$ denote the occupation of the transition metal and Si atoms with 0.5 $\leq$ z $\leq$ 1 and z/2 $\leq$ y $\leq$ z (see Ref. \cite{Takamura:2009ei} for details). In the case of a regular Heusler (for Co$_2$FeSi), $S_{\mathrm{D-B}}$ and $S_{\mathrm{D-A,C}}$ match the ordering parameters $S_{\mathrm{L2_1}}$ and $S_{\mathrm{B2}}$, respectively. Experimentally, the long-range ordering parameters $S$ are obtained by comparing experimental and theoretical diffraction peak intensity ratios. The crucial anomalous dispersion contributions \cite{Patterson63} are fully taken into account in our analysis.

The influence of the annealing temperature on the D-A,C ordering is shown in Figure \ref{a004} b). It increases with increasing annealing temperatures. Co$_2$FeSi and Fe$_{1.25}$Co$_{1.75}$Si reach an $S_{\mathrm{D-A,C}}$ of nearly unity at $T_\mathrm{a} = 500^\circ$C. This means that sublattices A and C are almost completely occupied by transition metal atoms. The ordering systematically decreases with increasing Fe concentration. 
The D-B and D-A,C ordering parameters are shown in Figure \ref{S_vs_x_graph} as a function of x for $T_\mathrm{a} = 500^\circ$C. Co$_2$FeSi shows overall the best atomic ordering. $S_{\mathrm{D-B}}$ drops with increasing Fe concentration, which indicates increasing Fe(B)-Si disorder as Fe$_2$CoSi is approached. For Fe$_2$CoSi, this translates to a Si occupancy of 6.5\% on the A,C sites each, and 21\% on the B sites. Accordingly, the D site has 66\% Si occupancy, where 25\% would mean completely random distribution.

\subsection{Transport properties}
\begin{figure}[!t]
\centering
\includegraphics[width=3.5in]{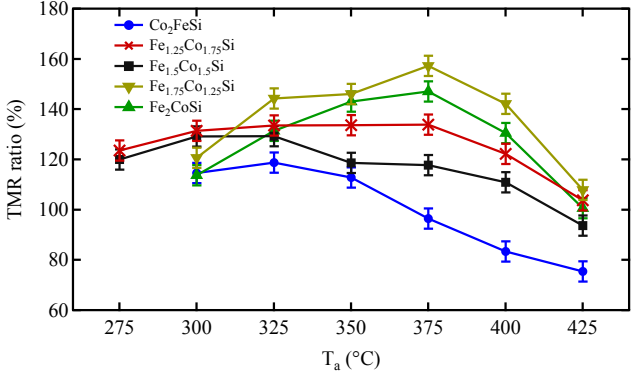}
\caption{Room temperature TMR ratio for various annealing temperatures.}
\label{TMRTa_graph}
\end{figure}

The influence of post-annealing on the TMR is studied with subsequent annealing of patterned samples in a magnetic field of 0.1\,T for 60\,min. The TMR ratio vs. annealing temperature for all stoichiometries is shown in Figure \ref{TMRTa_graph}. 
The TMR data were in all cases taken at a bias voltage of 10\,mV. The error bars are estimated from the standard deviation of the TMR values of at least ten elements. A maximum room temperature TMR ratio of 159\% was achieved for Fe$_{1.75}$Co$_{1.25}$Si at $T_\mathrm{a} = 375^\circ$C. 

The TMR vs. $T_\mathrm{a}$ curves are quite different for the various stoichiometries. For Fe$_2$CoSi and Fe$_{1.75}$Co$_{1.25}$Si, they are very similar and start with a TMR ratio of about $120\%$ at $300^\circ \mathrm{C}$ annealing temperature, which increases up to 159\% for Fe$_{1.75}$Co$_{1.25}$Si. The Fe$_2$CoSi compound also reaches the highest TMR ratio of 147\% at 375$^\circ$C post-annealing temperature. With further increased annealing temperature, the TMR values decrease to ratios below those at 300$^\circ$C. The increasing TMR ratio for temperatures $\leq$\,375$^\circ$C points to an improving spin polarization due to a better crystallinity of the Heusler electrode and perhaps to a more crystallized tunneling barrier.

The decreasing TMR ratio for annealing temperatures above $375^\circ$C is well known and arises from Mn interdiffusion out of the MnIr counter electrode into the CoFe electrode and its interface to the MgO barrier \cite{Hayakawa06}. The Mn destroys the magnetic and electronic structure of the CoFe, resulting in a decreasing TMR ratio.

A slightly different behavior can be found for the Fe$_{1.25}$Co$_{1.75}$Si tunneling element. The TMR ratio remains constant for 300$^\circ$C\,$\leq T_\mathrm{a}\leq$\,375$^\circ$C and starts to decrease at temperatures higher than 375$^\circ$C. The reason for this is again the Mn diffusion. The fact that the TMR ratio remains constant indicates that opposing effects may be present: although the D-A,C ordering parameter increases, the TMR ratio stays constant. It is also possible (but unlikely) that the crystallinity of the barrier does not change for these compositions, microstructural analysis would be required to assess this conjecture.

The tunnel junction based on the Co$_2$FeSi film exhibits different behavior: the maximum TMR ratio of 118\% is already reached at $T_\mathrm{a} = 325^\circ$C. For higher temperatures, the TMR ratio decreases, although the D-A,C ordering strongly increases between 325$^\circ$C and 375$^\circ$C. We can not explain this decrease with the Mn diffusion, which starts not before $T_\mathrm{a} = 375^\circ$C in the other samples. This suggests that the spin polarization \textit{reduces intrinsically as crystallinity is improved} (with respect to the D-A,C ordering). This finding agrees with earlier investigations of Co$_2$FeSi based MTJs with AlO$_x$ barriers, which have found a spin polarization that is considerably lower than that of Co$_2$MnSi \cite{Kubota09}.

\begin{figure}[!t]
\centering
\includegraphics[width=3.5in]{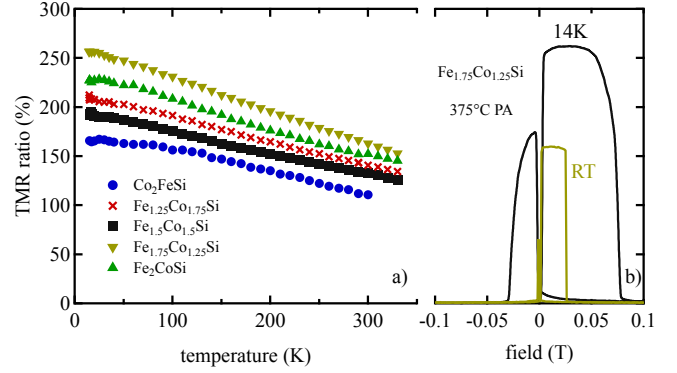}
\caption{a) Temperature dependent TMR ratio. The annealing temperatures are $325^\circ$C for Co$_2$FeSi, Fe$_{1.25}$Co$_{1.75}$Si and Fe$_{1.5}$Co$_{1.5}$Si and $375^\circ$C for Fe$_{1.75}$Co$_{1.25}$Si and Fe$_2$CoSi. b) Major loop of Fe$_{1.75}$Co$_{1.25}$Si at room temperature and 15\,K.}%
\label{TMR-T}%
\end{figure}

Temperature dependent measurements were carried out on MTJs with a size of (7.5$\times$7.5)\,$\mu$m$^2$ for all stoichiometries.  The annealing temperatures are 325$^\circ$C for Co$_2$FeSi, Fe$_{1.25}$Co$_{1.75}$Si and Fe$_{1.5}$Co$_{1.5}$Si and 375$^\circ$C for Fe$_{1.75}$Co$_{1.25}$Si and Fe$_2$CoSi, in order to observe the highest possible TMR ratio. The TMR versus temperature dependence is presented in Figure \ref{TMR-T} a). 
The MTJ based on the Fe$_{1.75}$Co$_{1.25}$Si Heusler film shows a TMR ratio of 262\% at 15\,K and 159\% at room temperature, respectively. Figure \ref{TMR-T} b) shows the corresponding major-loops. The scaling factor TMR(15\,K)/TMR(290\,K) is about $1.5$ for all samples.

Drewello \textit{et al.} reported TMR ratios of $76\%$ at room temperature and $134\%$ at 13\,K for Co$_2$FeSi based MTJs with MgO barrier and Co$_{70}$Fe$_{30}$ counter electrode \cite{Drewello:2012uha}. The enhanced TMR ratio in the present work is attributed to better film growth because of the Cr buffer. In Ref. \cite{Drewello:2012uha}, the ratio was TMR(15\,K)/TMR(290\,K)=1.7, which is close to the values observed here. To the best of our knowledge, we have observed the highest TMR obtained with a Co$_2$FeSi/MgO/CoFe junction. Higher ratios with a Co$_2$FeSi electrode have only been obtained with a CoFeB counter-electrode \cite{lim08}.

\begin{figure}[!t]
\centering
\includegraphics[width=3.5in]{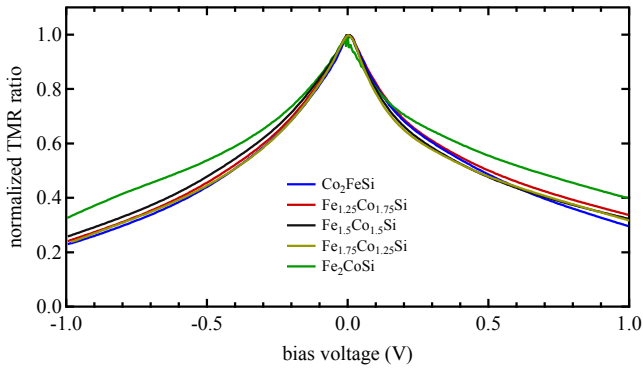}
\caption{TMR vs. voltage dependence measured at 15\,K.}%
\label{TMR_V_graph}%
\end{figure}
Figure \ref{TMR_V_graph} depicts the voltage dependence of the TMR for all stoichiometries measured at $15\,$K. The curves are normalized to the TMR at zero bias. 
With the exception of Fe$_2$CoSi the curves are similar to each other and resemble those obtained for many Heusler compounds. The voltage required to reduce the TMR to the half zero-bias value is between 400\,mV and 500\,mV. The result for Fe$_2$CoSi shows a slower decay of the TMR with increasing bias voltage, and a small bump at around -0.8\,V. The question if these features can be attributed to band structure effects is subject of ongoing work.
\section{Conclusions}
For the first time, magnetic tunnel junctions based on an inverse Heusler compound have been prepared. We have investigated the crystallographic and transport properties of Fe$_{1+x}$Co$_{2-x}$Si based magnetic tunnel junctions. High TMR ratios were obtained for the intermediate stoichiometry Fe$_{1.75}$Co$_{1.25}$Si. This can be attributed to an adjustment of the Fermi energy within a minority spin pseudogap. For this case, we have observed a correlation between the long-range ordering parameters and the TMR as a function of the annealing temperature. In the Co$_2$FeSi case, however, no such correlation was observed, which indicates a low intrinsic spin polarization, in agreement with earlier reports.

\section*{Acknowledgment}
The authors gratefully acknowledge financial support from Bundesministerium f\"ur Bildung und Forschung (BMBF) and Deutsche Forschungsgemeinschaft (DFG, contract nr. RE 1052/RE24-1).

\end{document}